\documentclass[preprint,onecolumn,superscriptaddress]{revtex4-1}

\usepackage{bm}

\usepackage[english]{babel}
\usepackage{amsmath}
\usepackage{amssymb}
\usepackage{dsfont}
\usepackage{color}
\usepackage{hyperref}
\usepackage{graphicx}
\usepackage{soul}
\usepackage{color}
\usepackage{braket}
\usepackage[normalem]{ulem}
\pdfoutput=1

\begin{document}

\title{Multiplexed Single-Mode Wavelength-to-Time Mapping of Multimode Light}

\author{Harikumar K. Chandrasekharan\footnote{These authors contributed equally}}
\affiliation{Scottish Universities Physics Alliance (SUPA), Institute of Photonics and Quantum Sciences, Heriot-Watt University, Edinburgh, EH14 4AS, UK}
\affiliation{Institute of Biological Chemistry, Biophysics and Bioengineering, Heriot-Watt University, Edinburgh, EH14 4AS, UK}
\author{Frauke Izdebski\textsuperscript{a}}
\affiliation{Scottish Universities Physics Alliance (SUPA), Institute of Photonics and Quantum Sciences, Heriot-Watt University, Edinburgh, EH14 4AS, UK}
\author{Itandehui Gris-S{\'a}nchez}
\affiliation{Department of Physics, University of Bath, Claverton Down, Bath, BA2 7AY, UK}
\author{Nikola Krstaji{\'c}}
\affiliation{Institute for Integrated Micro and Nano Systems, School of Engineering, University of Edinburgh, Edinburgh, EH9 3JL, UK}
\author{Richard Walker\footnote{Present address: Photon Force Ltd, Edinburgh, EH3 7HA}}
\affiliation{Institute for Integrated Micro and Nano Systems, School of Engineering, University of Edinburgh, Edinburgh, EH9 3JL, UK}
\author{Helen L. Bridle}
\affiliation{Institute of Biological Chemistry, Biophysics and Bioengineering, Heriot-Watt University, Edinburgh, EH14 4AS, UK}
\author{Paul A. Dalgarno},
\affiliation{Institute of Biological Chemistry, Biophysics and Bioengineering, Heriot-Watt University, Edinburgh, EH14 4AS, UK}
\author{William N. MacPherson}
\affiliation{Scottish Universities Physics Alliance (SUPA), Institute of Photonics and Quantum Sciences, Heriot-Watt University, Edinburgh, EH14 4AS, UK}
\author{Robert K. Henderson}
\affiliation{Institute for Integrated Micro and Nano Systems, School of Engineering, University of Edinburgh, Edinburgh, EH9 3JL, UK}
\author{Tim A. Birks}
\affiliation{Department of Physics, University of Bath, Claverton Down, Bath, BA2 7AY, UK}
\author{Robert R. Thomson}
\affiliation{Scottish Universities Physics Alliance (SUPA), Institute of Photonics and Quantum Sciences, Heriot-Watt University, Edinburgh, EH14 4AS, UK}

\begin{abstract} 
We demonstrate that photonic lanterns based on tapered multicore fibres provide an efficient way to couple multimode states of light to a two-dimensional array of Single-Photon Avalanche Detectors (SPADs), each of which has its own Time-to-Digital Converter (TDC) for Time-Correlated Single-Photon-Counting (TCSPC). Exploiting this capability, we demonstrate the multiplexed single-mode wavelength-to-time mapping of multimode states of light using a multicore-fibre photonic lantern with 121 single-mode cores, coupled in a one-to-one fashion to 121 SPADs on a 32 $\times$ 32 pixel CMOS SPAD array. The application of photonic lanterns for coupling multimode light to SPAD arrays in this manner may find wide-ranging applications in areas such as Raman spectroscopy, coherent LIDAR and quantum optics.

\end{abstract}

\maketitle

\section{Introduction}
In many photonic application areas, single-mode optical waveguides are preferred over multimode waveguides for the transmission of light. This is because single-mode systems do not suffer from either inter-modal dispersion or inter-modal coupling. These effects are highly detrimental in areas such as telecommunications (where they reduce the bandwidth-length product of a fibre-optic link) and fibre-optic imaging (where they scramble the information required to generate a clear image). Yet, there are also areas where multimode waveguides are preferred. For example, photon-starved applications such as astronomy \cite{Barden:1981hl}  almost solely use multimode fibres, since they enable the efficient collection and transmission of multimode signals to instruments for analysis.

The benefits of multimode waveguides, such as higher collection efficiencies and increased tolerances to misalignment, can be combined with the transmission and processing benefits of  single-mode optical waveguides by harnessing the efficient multimode-to-single-mode coupling capabilities offered by photonic lanterns \cite{Birks:2015eg,Hawthorn:2010sl,2005OptL...30.2545L,Birks:2012od,Thomson:2011ge,2010OExpr..18.4673N}. In its most general form, a photonic lantern is a gradual, and ideally adiabatic, transition between a multimode waveguide, supporting \textit{N} modes, and an array of \textit{N} single-mode waveguides. The multimode end of the photonic lantern facilitates the efficient collection of incoherent multimode states of light, while the single-mode cores enable the benefits of single-mode confinement and transmission to be harnessed. 

Photonic lanterns were originally developed for future applications in ground-based astronomy. There it is desireable to utilise complex single-mode Fibre-Bragg-Gratings (FBGs) for removing unwanted atmospheric lines from the celestial light of interest (OH-line suppression) \cite{Ellis:2011gk,Spaleniak:2013kl}, but it is also essential to efficiently collect the multimode light that forms the telescope point spread function \cite{Ozdur:2013ez,Betters:2013fv}. The applications of photonic lanterns are now moving well beyond astronomy. For example, photonic lanterns are attracting considerable interest in space-division-multiplexed telecommunications, where they enable the efficient coupling of light between multiple single-mode fibres and a few-mode fibre \cite{LeonSaval:2014cv,Yerolatsitis:14k,2014NaPho...8..865V}. They are also of interest for coherent LIDAR \cite{Ozdur:2015gs}, where they facilitate greatly-improved multimode signal collection along with excellent mode-matching to a single-mode local oscillator for clean heterodyne mixing.

In this paper, we demonstrate that multicore fibre (MCF) based photonic lanterns can be used to efficiently and adiabatically couple multimode states of light to a two-dimensional Single-Photon Avalanche Detector (SPAD) array (Megaframe), where each SPAD has its own Time-to-Digital Converter (TDC) for Time-Correlated Single-Photon-Counting (TCSPC) \cite{Richardson:2009ij,Gersbach:2012it,Nik:2015}. SPAD arrays such as the Megaframe have recently attracted considerable attention for applications in areas such as fluorescence lifetime \cite{Poland:2015} and light-in-flight \cite{Gariepy:2010} imaging. To achieve acceptable dark-count-rates, the photosensitive area of each SPAD on the Megaframe is only $\approx$ 6 $\mu$m in diameter. This, combined with the fact that each SPAD has its own dedicated TDC electronics, means that the SPADs are spaced on a 50 $\mu$m $\times$ 50 $\mu$m grid, and the SPAD array exhibits a physical fill factor of $\approx\!1$~$\!\%$. In this work, we demonstrate that the use of a photonic lantern for coupling light to the SPAD array increases the effective fill factor of a subset of SPADS on the array by a factor of $\approx\!$ 30, from 1~$\!\%$ to 30~$\!\%$, whilst also combining the benefits of multimode light collection and single-mode transportation and delivery. These capabilities may prove to be enabling in many single-photon application areas. For example, with specific reference to the coherent LIDAR application mentioned above, the use of MCF lanterns and multipixel SPAD arrays offers a route to high multiplex gains, with a consequent increase in sensitivity and measurement speed. More generally, the splitting and reformatting capability offered by photonic lanterns could significantly increase data acquisition rates in single-photon counting applications that suffer from pulse pile-up \cite{Harris:1979}, by efficiently coupling the input signal to many SPADs in a scalable manner. In more fundamental areas of physics, the use of photonic lanterns for adiabatically coupling multimode states of light to multipixel SPAD arrays may also prove to be powerful in quantum optics, for observing multi-photon, multimode interference in high-dimensional quantum systems \cite{Defienne:2015vc}.

\begin{figure}[t!]
\includegraphics[trim=.08cm .08cm .08cm .08cm, clip, width =15cm]{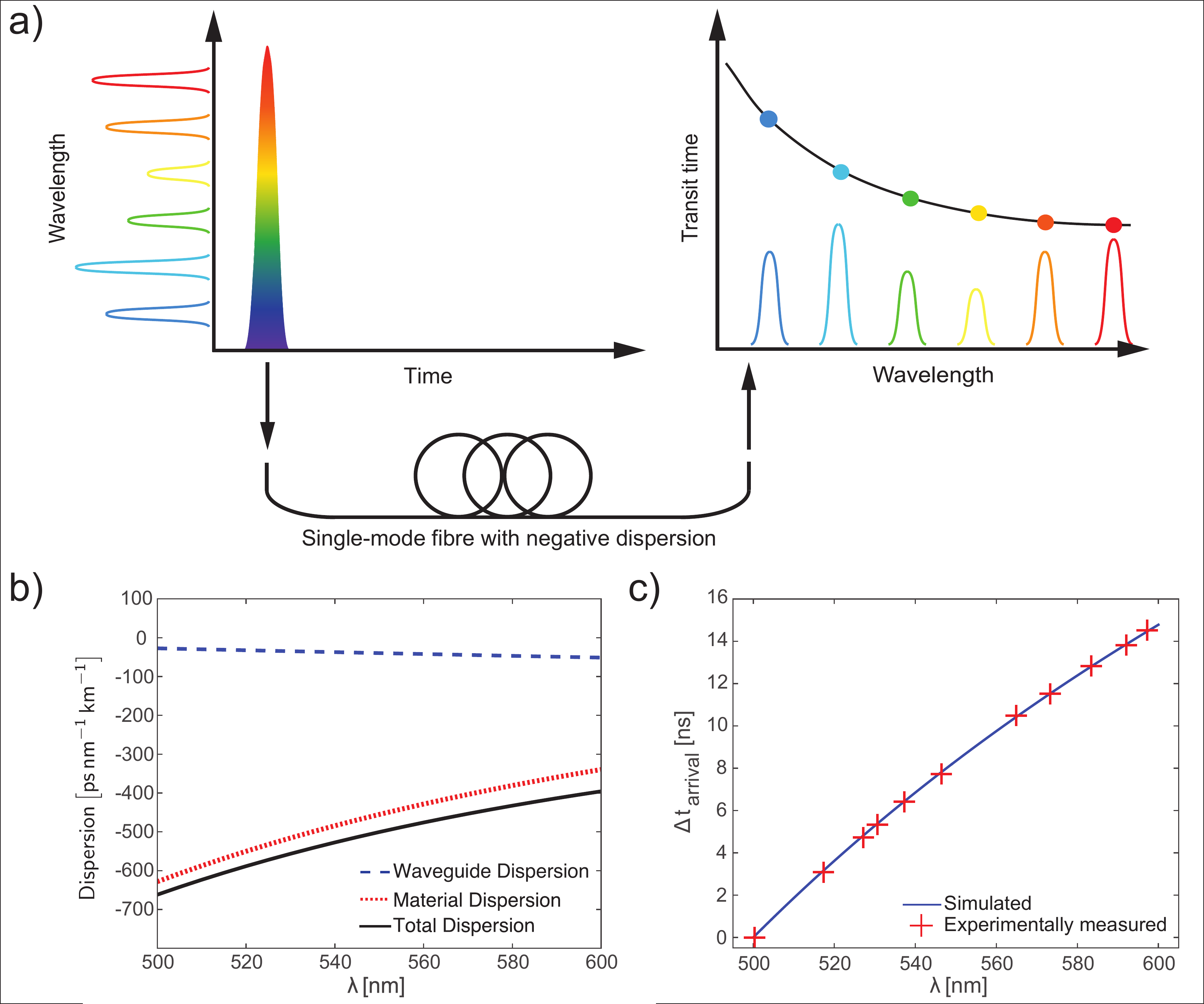}
\caption {a) Conceptual representation of the wavelength-to-time mapping phenomena in fibre-optics. On the left of the figure, a short broadband pulse is coupled into a long length of single-mode fibre. On the right of the figure, the different colours that form the input pulse emerge from the fibre at different times due to the chromatic dispersion of the fibre, resulting in the mapping of wavelength to arrival time. b) Simulated dispersion of the MCF cores. Also shown are the approximate contributions due to waveguide and material dispersion. c) Expected differences in arrival times $\Delta$$t_{arrival}$ after propagating along 290 m of the MCF, according to the dispersion profile shown in b). The crosses represent real data measured for the central core of the MCF. Note that larger values of $\Delta$$t_{arrival}$ indicate a shorter propagation time along the fibre.}
\label{Wavelength_Time}
\end{figure}

To showcase the enabling nature of the techniques we present here, we use a photonic lantern for multiplexed single-mode TCSPC-based Wavelength-to-Time Mapping (WTM) of multimode states of light. WTM is a phenomenon that occurs when a wave packet of light (or indeed any other type of wave) propagates through a sufficiently long length of a dispersive medium. As shown conceptually in  Figure~\ref{Wavelength_Time}a, WTM occurs in fibre optics when a short pulse of broadband light propagates along a length of dispersive optical fibre. The input pulse can be considered as being composed of an infinite set of pulses with different central wavelengths, each of which propagates at a different group velocity. As discussed in \cite{Meng:2015it}, after propagating down a length of fibre, two such pulses with different central wavelengths ($\lambda_1$ and $\lambda_2$) will become separated by a temporal duration ($\Delta t$), according to Equation \ref{Resolution}:

\begin{equation}
\Delta t = \int_{\lambda_1}^{\lambda_2} L  D (\lambda)d\lambda
\label{Resolution}
\end{equation}
where $L$ is the length of the single-mode fibre and $D$ is the group velocity dispersion (GVD) of the fibre.

WTM is an established technique in optics, where it has already found applications in real-time spectroscopy\cite{Whitten:1979bh,Chinowsky:1999hl,Solli:2008bg,Meng:2015it}. For example, WTM has provided significant insights into the nonlinear dynamics and wavelength correlations that occur during broadband supercontinuum generation, insights that would simply not be possible using measurements that are averaged over multiple pulses \cite{Wetzel:2012eyl}. WTM is now also beginning to find real-world applications in areas such as Raman spectroscopy where, in combination with TCSPC, it enables the acquisition of Raman spectra without the requirement for a conventional spectrometer \cite{Petrov:2012uy, Meng:2015it, 2015OExpr..23.5078T}. To date, however, all work on WTM based TCSPC Raman spectroscopy has been limited to the use of one single-photon detector, restricting the signal acquisition rates possible. In this paper, we demonstrate that MCF based photonic lanterns, combined with multi-pixel SPAD arrays, now offer a powerful route to overcome  these limitations, by enabling multiplexed single-mode TCSPC-WTM of multimode states of light.

\section{Experimental Details}
\subsection{Experimental Setup}
\begin{figure}[t!]
\centering
\includegraphics[trim=3.9cm 0cm 0cm 0cm, clip, width =15cm, angle = 0]{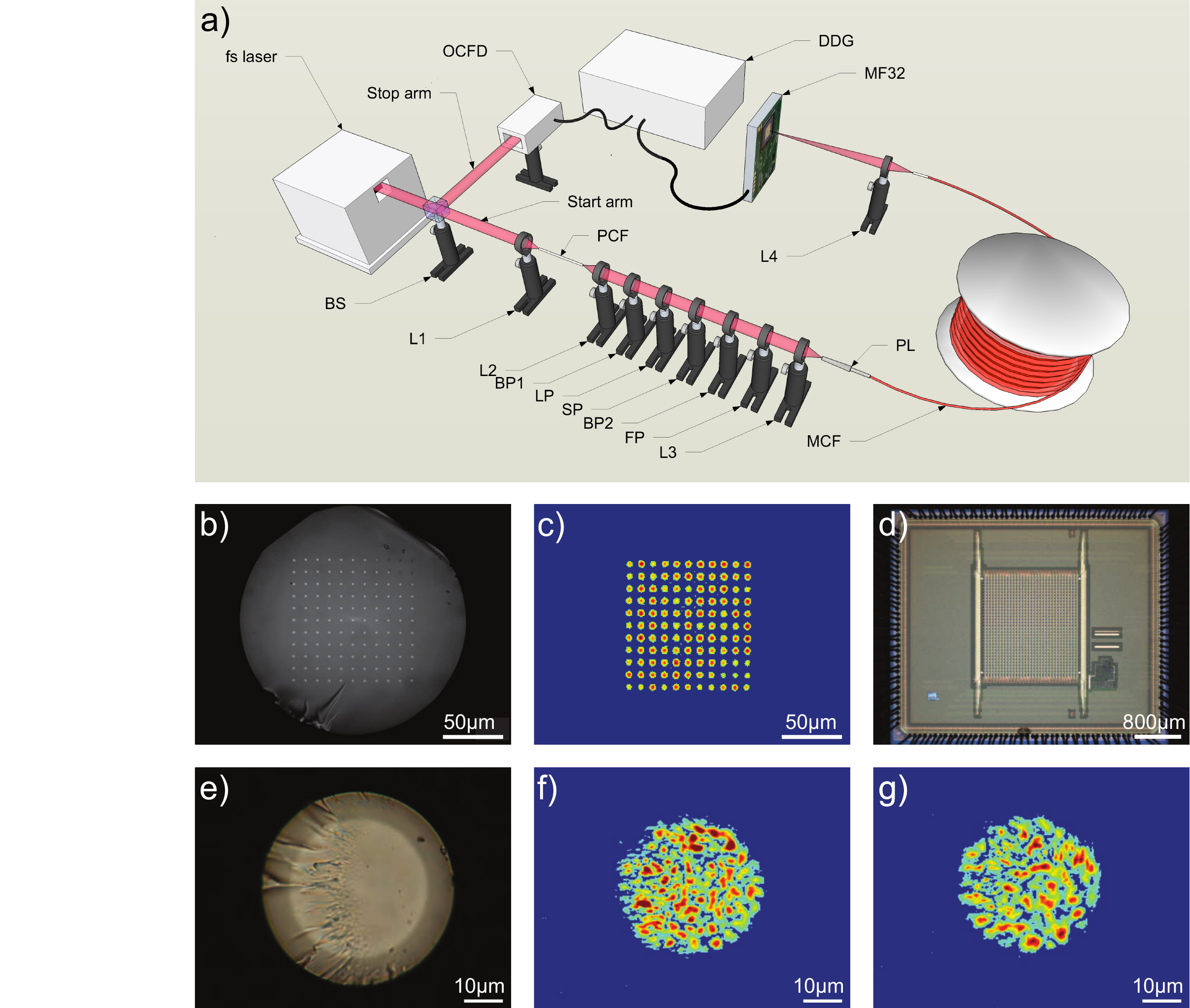}
\caption{a) Experimental setup used for multiplexed single-mode wavelength-to-time mapping. b) Micrograph of the MCF, showing the $11 \times 11$ square array of single-mode cores. c) False colour image of the output of a 9 m length of MCF when coupling 532 nm light into a photonic lantern at the opposite end. d) Micrograph of the Megaframe detector (MF32) with $32 \times 32$ SPADs. e) Micrograph of the multimode end of a photonic lantern. f) and g) False colour images of different output intensity profiles from (e) when exciting different single-mode MCF cores at the opposite end of the photonic lantern using 532 nm light.}
\label{Exp_Setup}
\end{figure}

The experimental setup for the WTM experiments, and the key components used in this setup, are presented in Figure~\ref{Exp_Setup}. For multiplexed single-mode WTM, we designed and fabricated a custom MCF. The MCF consists of a square array of $11 \times 11$ single-mode cores with a core-to-core pitch of  10~$\mu m$, as shown in Figure~\ref{Exp_Setup}b. The MCF cores have a diameter of 1.63 ~$\mu m$, and are formed from germanium doped silica. The cladding material is pure silica, and the core-cladding refractive index contrast is $\Delta n = 1.66 \times10^{-2}$. This design facilitates single-mode operation at wavelengths longer than  $\lambda =470$~nm, with negligible core-to-core coupling for  $\lambda<610$~nm after 300~m of propagation. Figure~\ref{Wavelength_Time}b presents the simulated GVD of the MCF cores throughout the 500~nm to 600~nm region,
where it can be seen that the total simulated GVD varies between -661.8~ps nm$^{-1}$ km$^{-1}$ and -395.6~ps nm$^{-1}$ km$^{-1}$ respectively. The GVD simulations were performed using the refractive index dispersion data for undoped and Ge-doped silica as detailed in \cite{Butov2002301} (samples 3 and 9 respectively in Table 1). Based on the simulated GVD, Figure~\ref{Wavelength_Time}c presents the predicted difference in arrival times for pulses with different central wavelengths after propagating along a 290~m length of the MCF.

As shown in Figure~\ref{Exp_Setup}a, the experimental setup for multiplexed WTM consists of two different arms for the TCSPC measurement. In the start arm, 1064~nm femtosecond laser pulses at a repetition rate of 500~kHz are focused into a 20~cm long Photonic Crystal Fibre (PCF) \cite{Stone08} using lens L1 to generate a broadband supercontinuum. The supercontnuum is collimated using lens L2 and projected through a series of spectral filters. Firstly, the pump beam is blocked and the supercontinuum narrowed using a bandpass filter (BP1) that passes only light within the 400~nm to 700~nm range. For all measurements, the wavelength range of the supercontinuum was further limited to between $\lambda_{min} =500$~nm and $\lambda_{max} =600$~nm using the longpass (LP) and shortpass (SP) filters. For calibration purposes, an angle-tuned Fabry-P{\'e}rot (FP) interference filter and bandpass filter (BP2) are also inserted into the beampath to produce spectrally narrow tuneable pulses of light. The filtered light is then focused using lens L3 into the multimode end of the photonic lantern (PL) \cite{Birks:2015eg}(Figure~\ref{Exp_Setup}e), where it excites a coherent multimode state (Figures 2f $\&$ g). The photonic lantern transition then adiabatically couples this state to the 121 single-mode cores of a 290~m long length of MCF. At the other end of the MCF, the single-mode cores are directly imaged onto the Megaframe (MF32) SPAD array (Figure~\ref{Exp_Setup}d) using lens L4, with the magnification and alignment carefully controlled to couple each MCF core to a separate SPAD. As an example, Figure~\ref{Exp_Setup}c presents the recorded output from a 9 m long MCF when light is coupled into the photonic lantern at the opposite end. The photonic lantern effectively increases the fill factor of a subset of SPADs on the Megaframe, by spatially rearranging the input light into a pattern which can then be efficiently coupled to the photosensitive areas of each pixel. As detailed in the Methods section, we evaluated the effective fill factor to be $\approx$ 30~$\!\%$, a significant increase on the $\approx$ 1~$\!\%$ physical fill factor of the Megaframe itself. For the stop arm of the setup, a small fraction of the fs-laser light is tapped using a beam splitter (BS) and coupled onto an Optical Constant Fraction Discriminator (OCFD). The OCFD generates a stable electrical stop pulse for the TCSPC.

The Megaframe is a Complementary Metal Oxide Semiconductor (CMOS) SPAD array with TCSPC capability, consisting of a square array of $32 \times 32$ pixels, each with a $\approx 6$~$\mu m$ diameter photosensitive area and a pixel pitch of 50~$\mu m$. 
As already stated, each pixel has its own TDC, enabling the array to generate a photon arrival timestamp with a time resolution of 53~ps. Each timestamp is 10 bits long, resulting in a dynamic range of 54~ns. The Megaframe is designed to operate in reversed start-stop TCSPC mode. Here the detection of a single photon at the Megaframe starts the TCSPC measurement, and the trigger pulse generated by the OCFD stops the measurement. An electronic time delay using a Digital Delay Generator (DDG) is used to match the propagation time of the 290~m long MCF in the start arm. For the current architecture, each pixel can deliver 500000 timestamps per second to the field programmable gate array \cite{Nik:2015}. Before conducting any WTM experiments, the Megaframe was characterised in terms of high dark count rate (DCR) pixels. It was found that $\approx$15~$\!\%$ of the pixels possess a high DCR, and that these pixels are randomly scattered across the array. These pixels were removed from any post-processing of the WTM data. The instrument response function (IRF) of the instrument defines the maximum achievable timing resolution. Across the Megaframe, the full-width-half-maximum (FWHM) of the IRF varied from 137~ps to 174~ps, a variation that originates from the fact that each pixel has its own TDC.  Another important characteristic of the Megaframe is how uniformly the IRFs for the SPADs align in time due to variations in the time delay of the stop signal. 
The standard deviation of the variation in timing response about the mean was measured to be $\pm$ 265~ps. 
This shift was taken into account during calibration measurements.

\subsection{Calibration of the wavelength-to-time-mapping process}
To calibrate the wavelength-to-time mapping properties of the MCF, spectrally narrow pulses of light were selected from the supercontinuum. This was achieved by inserting the bandpass (BP2) and Fabry-P{\'e}rot filters into the beampath. The Fabry-P{\'e}rot could be angle-tuned to control the transmission wavelength, and the bandpass of BP2 was chosen such that only the light within the central peak of the Fabry-P{\'e}rot was selected. Before each TCSPC measurement, a spectrum of the filtered light was obtained using a commercial spectrometer placed directly after the Fabry-P{\'e}rot. Initially, light at $\lambda = 532$~nm was coupled into the multimode end of the photonic lantern, which was attached to 290~m of MCF. The average propagation time along the MCF was measured to be $\approx 1.3$~$\mu s$, while Figure 3a presents the difference in arrival times $(\Delta t)$ recorded for the $11 \times 11$ array of MCF cores (black squares represent pixels with a high DCR). The maximum difference in arrival time was measured to be 1.4~ns, and it is clear that the group velocity varies across the fibre, exhibiting a maximum in the bottom left of the array and increasing progressively across the MCF to the opposite corner. The precise reason for this variation is currently not known, but such differences could arise during the MCF fabrication, or are more likely due to spooling-induced strain across the MCF. These results immediately indicate that optimal multiplexed WTM using an MCF requires independent detection and calibration for each core of the MCF.

The WTM mapping process was then calibrated for each MCF core by angle tuning the FP to generate narrow band pulses across the 500 nm to 600 nm range. Figure 3b presents the WTM calibration curves across this range for the MCF cores with the highest and lowest group velocities, and also the central MCF core - the core that should be least affected by spooling-induced strain. Also shown in Figure 3(b) are 4\textsuperscript{th} order polynomial fits to the data. The equations of these fits are used in the following section to convert arrival time into wavelength. The arrival times for the central core are also plotted in Figure 1c, where it can be seen that the WTM process in the central core is functioning in close agreement with the simulated dispersion profile of the MCF reported in Figure 1b.

\begin{figure}[t!]
\includegraphics[width = 15 cm]{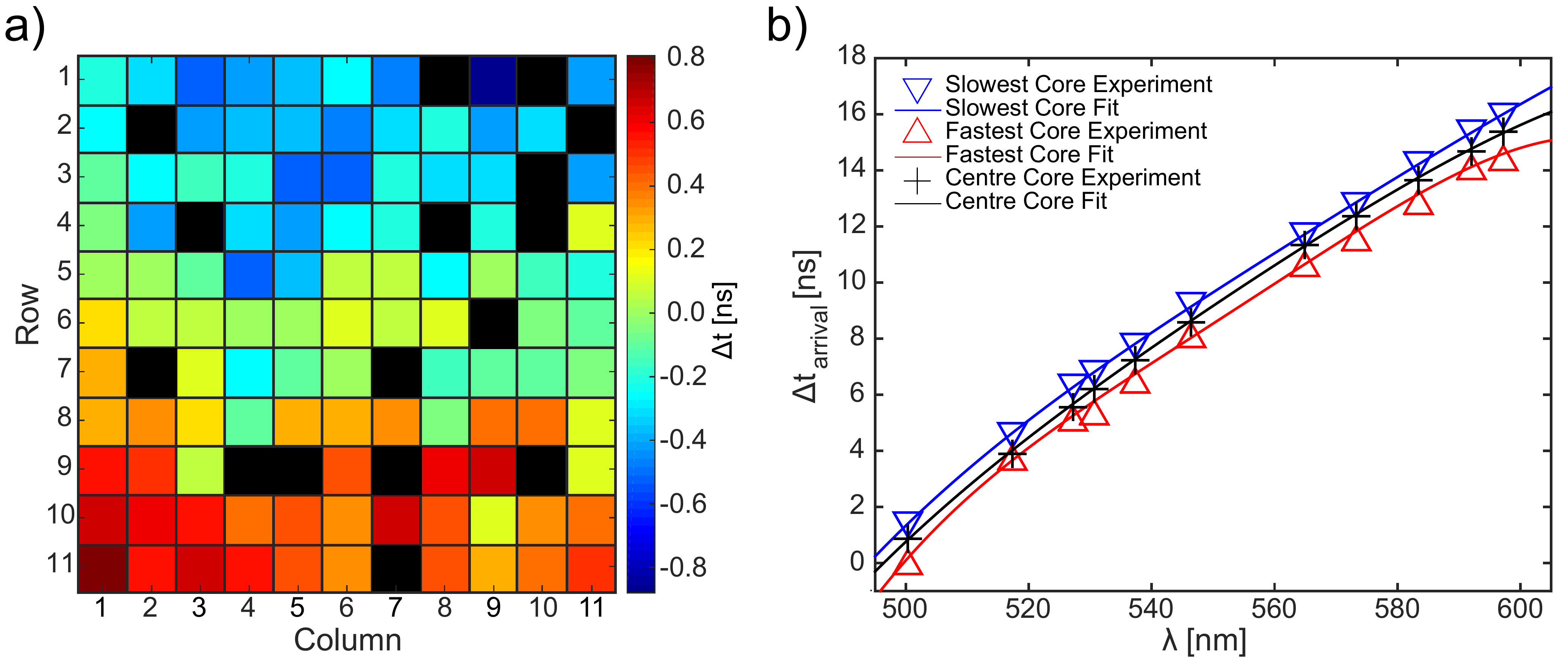}
\caption{Calibration of arrival times for each fibre core. a) Difference in arrival times $(\Delta t)$ for each core of the 290~m length of MCF measured at wavelength $\lambda =532$~nm. A black square indicates a pixel with a high DCR.  b) Measured (symbols) and fitted (solid lines) wavelength-dependent arrival times for the MCF cores with the highest and lowest group velocities, as well as the central MCF core.}
\label{Calibration}
\end{figure}

To probe the spectral resolution we can expect from our WTM system, we tuned the Fabry-P{\'e}rot to generate narrow band pulses at $\lambda =550.3$~nm. When measured using a spectrometer with a FWHM spectral resolution of 0.1~nm, the FWHM passband of the filter at this wavelength was measured to be 0.5~nm, while the WTM data obtained from the central MCF core indicated that the pulses exhibited a FWHM spectral width of 1.2~nm. The spectral resolution of the WTM spectra obtained from the central core was then estimated by convolving the filter passband spectrum with a Gaussian profile of increasing width, until the FWHM of the convolved data matched the 1.2~nm FWHM of the WTM spectrum. This occurred when the Gaussian FWHM was set to 0.96~nm, and this represents the approximate equivalent linewidth of the WTM spectra obtained from the central MCF core at 550~nm. The spectral resolution of any WTM spectra is, according to Equation~\ref{Resolution}, determined by the dispersion and length of the fibre, but also the IRF, which includes all instrument properties (the detector array, electronics, and the laser source jitter) that degrade the temporal precision of the measurement. We can calculate the expected resolution of a WTM spectrum at some wavelength by considering how widely spaced two pulses must be in wavelength if they are to be separated by the FWHM of the IRF ($\approx 150$~ps). Based on the arrival times for the central MCF core presented in Figure 3b, the dispersion is calculated to be $\approx-505.9$~ps nm$^{-1}$ km$^{-1}$ at 550.3~nm, and Equation~\ref{Resolution} indicates that the two pulses must be separated by 1.02~nm if they are to arrive separated by 150~ps. This value is close to the  0.96~nm resolution we evaluated experimentally using the Fabry-P{\'e}rot filter, and further confirms our WTM instrument is operating as would be expected according to theory. As shown in Figure 1b, the dispersion of the MCF cores is expected to increase (decrease) as we move to shorter (longer) wavelengths from 550 nm, and the central core is measured to exhibit a dispersion of -621.6~ps nm$^{-1}$ km$^{-1}$ and -417.8~ps nm$^{-1}$ km$^{-1}$ at 510 nm and 590 nm respectively. We would therefore reasonably expect the resolution of spectra obtained from this core to range from $\approx$ 0.8 nm and 1.2 nm according to Equation~\ref{Resolution}. In future applications, the final spectrum from the multiplexed WTM system will be obtained by summing the spectra obtained from each MCF core. In this case, the line function of the final instrument at some wavelength would be defined by the addition of the linefunctions exhibited by each core.

\begin{figure}[htbp]
\includegraphics[trim=0 0 0 0,clip, width=0.8\textwidth]{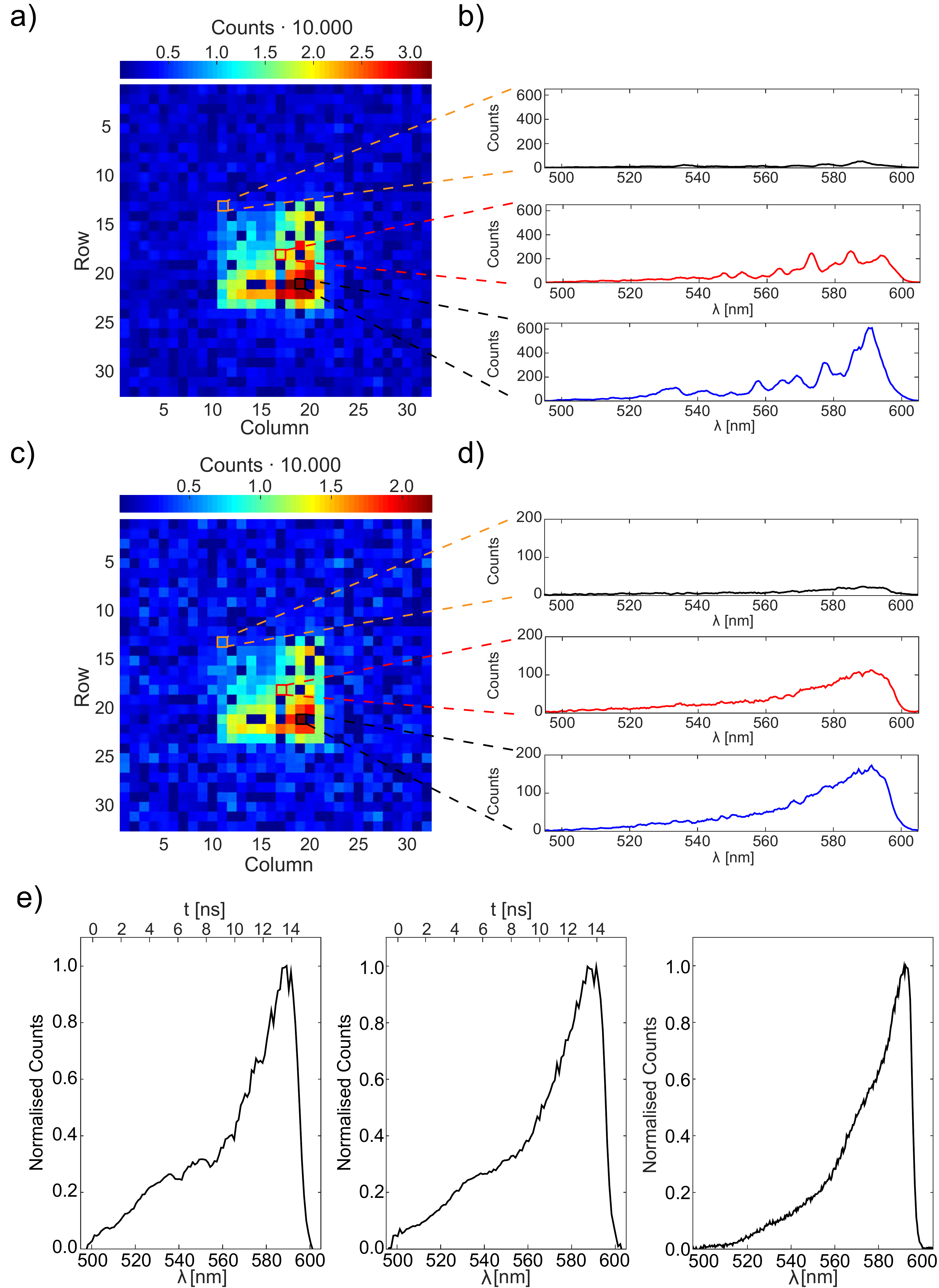}
\caption{Recorded broadband wavelength-to-time mapped spectra. a) Summed counts in each pixel of the Megaframe for the measurement without the rotating diffuser plate. b) Wavelength-to-time mapped spectra for three representative pixels. The spectra are different due to wavelength dependent coupling of the multimode light to the single-mode cores. c) Summed counts in each pixel for the measurement with the rotating diffuser plate. d) Wavelength-to-time mapped spectra for the same pixels selected in b) are shown, but this time all spectra are comparable due to the equalised excitation of all modes in the photonic lantern over the duration of the measurement. e) Summed spectra obtained from all illuminated pixels for the measurement without the rotating diffuser plate (left), with the rotating diffuser plate (middle), and a reference spectrum recorded with a conventional spectrometer (right). For the WTM spectra, the upper x-axis shows the difference in arrival times across the spectra.}
\label{Broadband}
\end{figure}

\subsection{Broadband multiplexed wavelength-to-time mapping}
For broadband WTM measurements, the Fabry-P{\'e}rot and bandpass (BP2) filters were removed, and the supercontinuum was spectrally narrowed to within the 500~nm to 600~nm range using only the BP1, SP and LP filters shown in Figure 2a. Broadband WTM spectra were then obtained in two ways. For the first set of measurements, the supercontinuum was focused directly into the multimode end of the photonic lantern, and Figure~\ref{Broadband}a presents a colourmap of the total counts for each pixel across the Megaframe, together with WTM spectra for three representative pixels, as shown in Figure~\ref{Broadband}b. When injecting the light into the photonic lantern in this manner, it was observed that each core at the end of the 290~m long MCF emitted light with very different spectra - the result of wavelength dependent coupling in the multimode end of the photonic lantern. It is also apparent in Figure~\ref{Broadband}a that the total counts across the $11 \times 11$ array varied significantly, and subsequent investigations revealed that this observation was primarily the result of variations in the propagation losses of the MCF cores.

The fact that the distribution of light across the MCF cores at the output of the fibre is strongly wavelength dependent immediately presents an interesting issue for the final objective of performing multiplexed WTM on arbitrary multimode states. Ideally, all MCF cores would exhibit the same propagation loss and the same coupling efficiency to the SPAD array, and all SPADs would exhibit the same detection efficiency.
But if this ideal situation is not achieved, there will be a wavelength-dependent loss of information in the final spectrum. Generally speaking, the importance of this will reduce as the number of modes in the measurement increases, since the fractional contribution to the final spectrum from any particular mode will reduce. To investigate how this loss of information has impacted the WTM spectra we measured, we performed a second set of WTM experiments, during which a rotating diffuser plate was placed directly in front of the multimode end of the photonic lantern. The effect of this diffuser plate was to excite all of the modes in the multimode end of the photonic lantern equally in a time-averaged manner, with the aim of exciting each single-mode in the MCF equally at all wavelengths. In this case, all MCF cores should produce the same WTM spectrum - the same spectrum we should measure without the diffuser plate if there is no loss of information. Figure~\ref{Broadband}c presents a colourmap of the summed counts obtained for each pixel across the Megaframe, while Figure~\ref{Broadband}d confirms that under these illumination conditions, all MCF cores produce very similar wavelength-to-time mapped spectra.

Figure~\ref{Broadband}e, presents the final normalised WTM spectra for the measurements taken without and with the rotating diffuser plate, obtained by adding together the individual WTM spectra obtained from each MCF core, together with a spectrum of the MCF output obtained using a conventional spectrometer. As can be seen, the spectrum obtained without the diffuser plate is close to that obtained with the diffuser plate, but there are some additional features in the spectrum obtained without the diffuser plate due to loss of some spectral information. Both WTM spectra are also in general agreement with the spectrum measured using a conventional spectrometer, although there is a clear increase in signal at shorter wavelengths in the WTM spectra compared to the conventional spectrum. The precise reason for this is yet to be confirmed, but could be the result of a wavelength dependent mode field diameter in the MCF, with shorter wavelengths being more confined to the core than longer wavelengths. In this case, since the SPADs are overfilled in our experiment, shorter wavelengths would be more efficiently coupled to the SPAD array than longer wavelengths. We also note that the photon detection efficiency of the Megaframe SPADs vary by $\approx$$ \pm$15~$\!\%$ of the mean across the 500 nm to 600 nm range \cite{Richardson2009}. Such variations would further contribute to the differences between the WTM spectra and the spectrum obtained with a conventional spectrometer, and would have to be taken into account in future applications.

\section{Conclusion}
We have demonstrated a new route to enable the efficient and adiabatic coupling of multimode states of light to multipixel SPAD arrays using MCF-based photonic lanterns. This approach is shown to increase the effective fill factor for a subset of SPADs withn the SPAD array by a factor of $\approx30$ when measured at 532 nm, from$\approx$ 1~$\!\%$ to  $\approx$ 30~$\!\%$, whilst also combining the benefits of multimode collection and single-mode transportation and delivery. Using this capability, we demonstrated the multiplexed single-mode WTM of multimode states of light. We showed that WTM spectra measured without the rotating diffuser plate are broadly in line with that recorded using a commercial spectrometer, but that they also exhibit some artifacts due to bad pixels on the Megaframe and variations in the propagation losses of the MCF cores. We also demonstrated that these articfacts can be removed by using a rotating diffuser plate to equally excite the modes in the photonic lantern in a time-averaged manner. For future applications, the scrambling function of the rotating diffuser plate could be replicated in a more efficient manner, by integrating multiple photonic lantern transitions into the MCF and gently agitating the fibre. Such an approach has already been demonstrated to result in excellent mode scrambling \cite{Birks:2012od}. We envisage that following the development of optimised MCF's and SPAD arrays, the techniques we have proposed and demonstrated here will open up new applications in photon-starved applications, such as Raman spectroscopy, coherent LIDAR and quantum optics.
\newpage

\section{Methods}
\subsection{Multi-Core Fibre}
For our experiments, it was of particular importance to design a fibre that exhibited negligible cross coupling between adjacent cores after propagating along a 290 m length of fibre, as this would compromise any temporal information obtained in the WTM experiment. To verify that cross coupling was negligible for wavelengths below $\lambda =610$~nm, light at different wavelengths was coupled into a single core of a 290 m length of the MCF and the distribution of light at the output of the fibre was measured with a CCD camera (Thorlabs DCC1645C).  No significant cross coupling for wavelengths below $\lambda =610$~nm was observed. The loss of the MCF was also investigated using the fibre cut-back method, and losses of the order of 0.2~dB m$^{-1}$ were inferred. 

\subsection{Photonic Lantern Fabrication}
To fabricate the photonic lantern, the MCF was threaded into a fluorine-doped silica capillary, which has a lower refractive index compared to the pure silica cladding of the MCF. The capillary was collapsed, by surface tension, on top of the MCF using an oxybutane flame. The cladded structure was then softened, using a similar flame, and stretched by a tapering rig, forming a biconical fibre-like structure. Finally, the multimode port of the photonic lantern was revealed by cleaving the centre of the tapered waist.

\subsection{Fill-Factor Enhancement}
The photonic lantern enables the efficient reformatting of multimode light into an array of single-modes that can be efficiently coupled to the SPADs on the Megaframe. This effectively increases the fill factor of a subset of SPADs on the Megaframe, and the following steps were taken to investigate this. First, 532 nm light was injected into the multimode end of a photonic lantern, which was attached to a 9 m length of MCF. A single-mode at the output of the MCF was then imaged onto the Megaframe with a magnification of $M = 1.5$. Under this magnification, the image of the core on the Megaframe is 2.55~$\mu m$ in diameter and it is reasonable to assume that all of the light falls inside the $\approx$ 6~$\mu m$ diameter photonsensitive area of each SPAD. The total number of counts over a set time period was determined for this core-SPAD combination and imaging magnification - this represents the maximum counts possible with a fill factor of 100~$\%$. A second measurement was then performed using a magnification of $M = 5$, the same magnification required to simultaneously couple the $121$ MCF-cores to $121$ SPADs. Under this magnification, the core image on the SPAD is 8.15~$\mu m$ in diameter, and the overlap of the mode on the 6~$\mu m$ diameter active area of the SPAD will be significantly reduced. Again, the number of counts obtained over a set time period was measured. The effective fill factor increase using the photonic lantern was determined by dividing the number of counts measured in the first measurement by the number of counts measured in the second measurement. Both measurements were repeated 4 times. In this manner, the effective fill-factor for a subset of SPADs on the SPAD array was found to be $30~ \% \pm 0.64 ~\%$ when using the photonic lantern, a significant improvement compared to the  1~$\!\%$ fill-factor without the photonic lantern.
    
\subsection{TCSPC Experimental Details}
The TCSPC experimental setup is shown in Figure~\ref{Exp_Setup}a. The TCSPC timer on the SPAD array is designed to work in reversed start-stop mode, where the detection of a single-photon at the SPAD array starts the timer and the detection of a trigger pulse stops the measurement. The setup therefore consists of start and stop arms. The start arm contains the photonic lantern and the 290 m long MCF. The optical signal for the start arm is obtained using a femtosecond laser (Fianium HE-1060-1~$\mu$J-fs), operating at a pulse repetition rate of 500~kHz and emission wavelength of 1064~nm, to generate a broadband supercontinuum by pumping a 20~cm long length of photonic crystal fibre \cite{Stone08} (Uni. of Bath). The supercontinuum pulses are spectrally filtered using bandpass, shortpass and longpass filters (Thorlabs - FEL0500, FES0600, and FESH0700). For callibration of the WTM process, narrowband pulses could be further selected by inserting a high order 550~nm Fabry-P{\'e}rot interference filter (developed by Edinburgh Biosciences and Delta Optical Thin Films) into the beam path, together with bandpass filters from Thorlabs (FWHM of 10~nm). The pulses in the start arm were focused onto the multimode end of the photonic lantern. At the output of the MCF, an achromatic lens was used to image the $121$ MCF single-mode cores onto an 11 $\times$ 11 array subset of pixels on the Megaframe SPAD array (U. of Edinburgh). A conventional spectrometer (Ocean Optics - USB2000+) was used to record the central wavelength of the pulses passed by the Fabry-P{\'e}rot interference filter. For the broadband measurements with the diffuser plate, the diffuser plate was rotated at 20~Hz using a chopper motor. The stop arm of the setup consists of an optical constant fraction discriminator (OCF401 - Becker $\&$ Hickl) which detects the arrival of a pump laser pulse and converts it into a stable electrical trigger signal to stop the TCSPC measurement. In order to match the time delay of 1.33~$\mu s$ introduced by the 290~m long fibre in the start arm, a digital delay generator (DG645 - Stanford Research Systems) was used to electronically delay the output pulse from the OCFD. In our measurements, each SPAD used in the experiment detected up to 0.0002 photons per pulse. All measurements were therefore obtained in the photon-starved regime required for TCSPC. Acquiring the spectra took 480 s, due to the low pulse repetition rate of the laser $f_{Laser}= 500$~kHz.

\section{Acknowledgements}
R.R.T acknowledges funding from the STFC in the form of an STFC Advanced Fellowship (ST/H005595/1). R.R.T, T.A.B., R.K.H, W.N.M, P.A.D, and H.L.B acknowledge funding through the STFC-CLASP scheme (ST/K006509/1, ST/K006460/1, ST/K006584/1). N.K. was suported via the EPSRC-IRC (EP/K03197X/1). We thank ST Microelectronics, Imaging Division, Edinburgh, for their support in the manufacture of the Megaframe chip. The Megaframe project has been supported by the European Community within the Sixth Framework Programme IST FET Open. R.R.T thanks Renishaw plc for support through the Renishaw-Heriot-Watt Strategic Alliance. R.R.T and H.K.C thank Heriot-Watt University for a James-Watt Studentship. We thank Dr Jim Stone (U. of Bath) for providing the supercontinuum fibre. We thank Prof. Des Smith FRS for providing us with the Fabry-P{\'e}rot multilayer filter - developed through the Eurostars Optitune Grant with partners Delta Optical Thin Films A/S and Edinburgh Biosciences Ltd.

\section{Contributions}
H.K.C and F.I. built and characterised the multiplexed wavelength-to-time mapping instrument. H.K.C performed the calibration and broadband measurements, F.I. designed the analysis programme and analysed the data. R.K.H. designed the CMOS SPAD pixel architecture, N.K. and R.W provided CMOS SPAD support, while N.K wrote the front end Labview software interfacing to Megaframe. The multicore fibre and photonic lantern were simulated by T.A.B., designed by I.G.-S. and T.A.B., and fabricated by I.G.-S.. H.L.B, P.A.D., and W.N.M. contributed to the discussion of results and the project motivation. F.I., H.K.C and R.R.T drafted the paper, and all authors discussed the results and contributed to the writing of the manuscript. R.R.T. conceived and led the project.

\section{Data}
Supporting data will be made available through the Heriot-Watt University Research Gateway Data Repository.

\section{Competing financial interests}
The authors declare no competing financial interests.
\section{Corresponding author}
Correspondence to 
R.R.Thomson@hw.ac.uk

\end{document}